\documentclass[aps,prb,twocolumn,preprintnumbers,amsmath,amssymb,superscriptaddress,floatfix]{revtex4}
\usepackage[dvips]{graphicx}
\usepackage{mathptmx}
\usepackage{verbatim}

\begin{document}
\title{Enhancement of non-local exchange near isolated band-crossings in graphene}
\author{Jeil Jung} \email{jeil@physics.utexas.edu}
\affiliation{Department of Physics, University of Texas at Austin, USA}
\author{Allan H. MacDonald}
\affiliation{Department of Physics, University of Texas at Austin, USA}
\date{\today{}}
\begin{abstract}
The physics of non-local exchange interactions in graphene sheets is studied within 
a $\pi$-orbital tight-binding model using a Hartree-Fock approximation and 
Coulomb interactions modified at short distances by lattice effects and at large distances
by dielectric screening.  We use this study to comment on the 
strong non-locality of exchange effects in systems with isolated band-crossings 
at energies close to the Fermi energy. 
We also discuss the role of lattice scale details of the effective Coulomb 
interaction in determining whether or not broken symmetry states appear at 
strong interaction strengths, and in determining the character of those states when 
they do appear. 
\end{abstract}


\maketitle

\section{Introduction}
Graphene sheets are ideal sp$^2$-hybridized pure carbon networks, 
and have attracted attention in recent years because of their appealing combination of 
theoretical simplicity and exceptional physical properties.\cite{novoselov,pkim,graphenereviews,castroneto}
Most electronic properties of graphene that have been studied experimentally can be successfully 
described in a non-interacting electron picture.\cite{castroneto}
Electron interaction effects are nevertheless clearly manifested 
in perpendicular magnetic fields where they lead to quantum Hall ferromagnetism \cite{qhallgraphene} and to the fractional 
Hall effect,\cite{andrei,bolotin,feldman}
when two\cite{bilayerinstb,bilayerinstb1} or more\cite{trilayerinstb} layers are stacked in a way which leads to  
flat bands near the Dirac point, 
and when ribbons with zigzag edges\cite{zigzag} are formed.
The strongest interaction effects so far observed in single-layer graphene in the absence 
of a magnetic field is the 
logarithmic velocity correction   
\cite{abrikosov,logdivref,logdivref1,logdivref2}
at momenta near the Dirac point, 
now apparent in photoemission measurements \cite{rotenberg}
and cyclotron mass measurements in suspended graphene. \cite{geim}
Recent Monte Carlo simulations of zero-field graphene suggest the more interesting possibility of a gap opening 
\cite{drutlahde} at the Dirac point for sufficiently strong interactions,
a property that would drastically modify electronic properties. 
Spontaneous gaps have still not been detected in single-layer samples, even when suspended\cite{bolotin} 
to reduce disorder and dielectric screening and,
although their appearance cannot be fully ruled out for cleaner suspended samples 
which might become available in the future, likely do not occur.

In this paper we use a $\pi$-band lattice-model Hartree-Fock calculation to show explicitly that the 
logarithmic velocity enhancement is related to non-local exchange interactions with power-law tails.
Our calculations provide a numerical estimate of the cut-off length which appears in the argument of 
the logarithm in the velocity enhancement expression and cannot be obtained from continuum model calculations.     
We also use our calculation to study the role that lattice scale physics plays in controlling whether or not gapped 
states can occur in single-layer graphene.  We show that the appearance of gapped states
is sensitive to the long-range of the Coulomb interaction.   
By solving self-consistent $\pi$-orbital Hartree-Fock equations,
we can assess the possibility of realizing topologically non-trivial states like those  
discussed by Raghu {\em et al.},\cite{zhang} who study an extended Hubbard 
model with next neighbor interactions.  

The paper is organized as follows.
We start in section II by briefly explaining our implementation of Hartree-Fock theory
for a $\pi$-orbital lattice model.  Here we define our model Hamiltonian, 
comment on how we handle complications due to the long-range of the Coulomb interaction, 
and discuss some other technical details of our calculations.
In section III we carry out a detailed study of the power-law non-local exchange interactions
and the logarithmic velocity enhancements they produce.
In section IV we present a mean-field phase diagram which identifies a 
variety of distinct broken symmetry solutions and captures the dependence of the 
competition between them on model parameters.
Finally we close the paper in section V with a discussion of our findings and of 
the general importance of highly non-local exchange interactions in
semi-metals or semiconductors with isolated band crossings, or weakly avoided crossings,
close to the Fermi level.

\section{$\pi$-orbital Hartree-Fock approximation}
The simplest tight-binding model for a carbon lattice
retains one atomic $2 p_z$ orbital on each lattice site and 
couples them with nearest
neighbor $pp\pi$ hopping\cite{slater} parameters.  
We use the conventions of Ref. [\onlinecite{hongki}], choosing
 a coordinate system in which the honeycomb's Bravais lattice has primitive vectors
\begin{equation}
\vec{a}_1=a(1,0) \ , \qquad\qquad \vec{a}_2=a\Big({1 \over 2},{\sqrt{3} \over 2}\Big),
\end{equation}
where $a = 2.46 {\rm \AA}$ is the lattice constant of graphene.
The reciprocal lattice vectors are then
\begin{equation}
\vec{b}_1={4\pi \over \sqrt{3}a}\Big({\sqrt{3}\over 2},-{1\over 2}\Big) \ , \quad \vec{b}_2={4\pi \over \sqrt{3}a}(0,1).
\end{equation}
Because nearest-neighbor hopping connects the honeycomb's two triangular sublattices,
the $2 \times 2$ tight-binding band Hamiltonian is purely off-diagonal:
\begin{equation}
\label{hamil}
{H}_0 \left( {\bf k} \right) = 
\begin{pmatrix}
     0 & \gamma_0 f\left( {\bf k} \right)   \\
\gamma_0 f^* \left( {\bf k} \right)   & 0  \\
\end{pmatrix}
\end{equation}
where $\gamma_0 = -2.6 eV$ is the hopping parameter and the on-site energy has been set to zero.
The factor
\begin{eqnarray}
f\left( {\bf k} \right) &=&   e^{ i k_y a / \sqrt{3} } \left( 1 + 2 e^{-i 3 k_{y} a / 2\sqrt{3}}  
                                       \cos \left(  \frac{k_x a}{2} \right)    \right)   
\end{eqnarray}
arises from the phase factors of the Bloch wavefunctions on neighboring sites.
We have neglected remote neighbor hopping which gives 
gives rise to electron-hole asymmetry, {\em i.e.} to $\bf{k}$-dependence of the 
sum of valence and conduction band energies. 
The convention for Bloch basis state phase factors which leads to this form of the 
Hamiltonian is  
\begin{equation}
\label{basis}
\left< {\bf r} | k \lambda \right> = \psi_{{\bf k} \lambda} \left({\mathbf r} \right)
= \frac{1}{\sqrt{N_K}} \sum_{i} e^{i {\mathbf k} \left( {\mathbf R}_i + {\bf \tau}_{l} \right)}
\phi \left({\mathbf r} - {\mathbf R}_i - {\bf \tau}_{l} \right) \eta_{\sigma}
\end{equation}
where $\eta_{\sigma}$ is the spin part of the wavefucntion, ${\bf \tau}_{l}$  is the
position of sublattice $l$ in the unit cell,
and $N_{K}$ is the number of unit cells in the systemt.
The label $\lambda = \left( l, \sigma \right)$ combines the lattice
site label $l$ and the spin label $\sigma$.

In this basis the Hartree-Fock Hamiltonian is 
\begin{widetext}
\begin{eqnarray}
\label{hfgen}
V_{HF} &=& \sum_{{\bf k} \lambda \lambda^{\prime}} U_H^{\lambda \lambda^{\prime}}
\left[ \sum_{{\bf k}^{\prime}}
\left<  c^{\dag}_{{\bf k}^{\prime} \lambda^{\prime}} c_{{\bf k}^{\prime} \lambda^{\prime}} \right>  \right]
c^{\dag}_{{\bf k} \lambda} c_{{\bf k} \lambda}    
- \sum_{{\bf k}^{\prime}\lambda \lambda^{\prime}} U_{X}^{\lambda \lambda'}
\left({\bf k}^{\prime} - {\bf k} \right)
\left<  c^{\dag}_{{\bf k}^{\prime} \lambda^{\prime}} c_{{\bf k}^{\prime} \lambda} \right>
c^{\dag}_{{\bf k} \lambda} c_{{\bf k} \lambda^{\prime}} 
\end{eqnarray}
where (dropping the spin index for simplicity)
\begin{eqnarray}
\label{hfgenh}
U_H^{\lambda \lambda^{\prime}} &=&
\left< {\bf k} \lambda {\bf k}^{\prime} \lambda^{\prime} \left| V \right| {\bf k} \lambda {\bf k}^{\prime} \lambda^{\prime} \right>  
= \int d {\bf r}_1 d {\bf r}_2 \left| \psi_{k \lambda} \left( {\bf r}_1 \right) \right|^2   V\left( \left| {\bf r}_1 - {\bf r}_2 \right| \right) 
\left| \psi_{{\bf k}^{\prime} \lambda^{\prime}} \left( {\bf r}_2 \right) \right|^2  \left( {\bf r}_2 \right)    \nonumber   \\
\label{uhar}
&=& \frac{1}{N_K^2} \sum_{i,j}^{N_K}
\int d{\bf r}_1 d{\bf r}_2   
 \left| \phi \left({\bf r}_1 - {\bf R}_i - {\bf \tau}_{l}  \right) \right|^2
 V\left( \left| {\bf r}_1 - {\bf r}_2 \right| \right)
 \left| \phi \left({\bf r}_2 - {\bf R}_j - {\bf \tau}_{l'}  \right) \right|^2  \\ 
\label{uex}
U_X^{\lambda \lambda'} \left( {\bf q} \right) &=&
\left< {\bf k} \lambda {\bf k}^{\prime} \lambda^{\prime} \left| V \right| {\bf k}^{\prime} \lambda {\bf k} \lambda^{\prime} \right>  
= \int d {\bf r}_1  d {\bf r}_2 \psi_{{\bf k} \lambda}^{*} \left( {\bf r}_1 \right) \psi_{{\bf k}^{\prime} \lambda}\left( {\bf r}_1 \right) 
 V\left( \left| {\bf r}_1 - {\bf r}_2 \right| \right)    
\psi^{*}_{{\bf k}^{\prime} \lambda} \left( {\bf r}_2 \right) \psi_{{\bf k} \lambda^{\prime}} \left( {\bf r}_2 \right)    \nonumber  \\
&=& \frac{1}{N_K^2}  \sum_{i^{\prime} j^{\prime}}
e^{i  \left(  {\bf k}^{\prime} - {\bf k} \right)    \left( {\bf R}_{i^{\prime}} + {\bf \tau}_{l} - \left( {\bf R}_{j^{\prime}} + {\bf \tau}_{l^{\prime}} \right) \right) }
V \left( \left| {\bf R}_{i^{\prime}} + {\bf \tau}_l  - {\bf R}_{j^{\prime}} - {\bf \tau}_{l^{\prime}}  \right| \right)       
\end{eqnarray}
\end{widetext}

We can simplify the two-body Coulomb integrals in Eqs. (\ref{uhar}) and (\ref{uex}) by combining the momentum-space  
representation for the Coulomb interaction (${\widetilde V}^{l l^{\prime}} \left( q \right) = {\widetilde V} \left( q \right) = 2 \pi e^2 / \varepsilon_r q$)
with the atomic orbital form factor 
$f({\bf q}) = \int d{\bf r} \, e^{- {\bf q} {\bf r}}  \left| \phi \left( {\bf r} \right) \right|^2$.
We use the explicit form
\begin{equation}
f\left( q \right) = (1 -  \left(r_o q\right)^2 ) / (  (1 + \left(r_o q \right)^2 )^4  )
\label{formfactor}
\end{equation}
obtained by Fourier transforming the radial charge distribution of a hydrogenic 
$2 p$ atomic orbital:
\begin{eqnarray}
\phi \left( r \right) = \frac{1}{\sqrt{4\pi}}   \frac{1}{\sqrt{24} \,\widetilde{a}_o^{3/2}}  \frac{r}{\widetilde{a}_o} e^{-r / 2 \widetilde{a}_o}.
\end{eqnarray}
The choice $\widetilde{a}_o = a_o/\sqrt{30} \AA$ reproduces the
the covalent bond radius of carbon $a_o = a/\left(2 \sqrt{3}\right)$.
Calculations in bilayer graphene suggest that a larger effective 
radius $\widetilde{a}_0 = 3 a_0 / \sqrt{30}$ is a better choice\cite{bilayerinstb1}
because it accounts crudely for $sp_{2}$ bonding orbital polarization.  
The two-body Coulomb integrals are then given by 
\begin{eqnarray}
U_H^{l l^{\prime}} &=&  \frac{1}{A} \sum_{\bf G} e^{ i {\bf G} \cdot \left(\tau_{l} - \tau_{l^{\prime}} \right) } 
\left| f\left( \left| {\bf G}   \right|\right) \right|^2 \,\, \widetilde{V} \left( \left|  {\bf G}   \right|  \right)  \label{mom1}
\\
U_X^{l \, l^{\prime}} \left( {\bf q} \right)    
&=& \frac{1}{A} \sum_{\bf G} \; e^{  i {\bf G} \cdot \left(\tau_{l} - \tau_{l^{\prime}} \right) } 
\left| f\left( \left| {\bf q}  -  {\bf G}  \right|\right) \right|^2 \,\, \widetilde{V}
\left( \left|  {\bf q} - {\bf G}  \right|  \right). \quad \quad   \label{mom2}
\end{eqnarray}
where ${\bf G}$ are the reciprocal lattice vectors and $A = N_K \, A_0$ is the system area.

We will also find it useful to consider an alternate model for interactions which assigns 
a value, $V_{\rm eff}(r)$  to the interaction strength between electrons which depends only on the 
distance between the lattice sites on which they reside.  
When expressed in terms of $V_{\rm eff}(r)$, 
\begin{eqnarray}
U_{H}^{l l^{\prime}} &\simeq& \frac{1}{N_K^2}
\sum_{i,j}  V_{\rm eff} \left(  \left| {\bf L}^{l l^{\prime}}_{ij} \right| \right)     \label{eq1}
\\
U_{X}^{l l^{\prime}} &\simeq& \frac{1}{N_K^2} \sum_{ij}^{N_K}
e^{i \left({\bf k}' - {\bf k} \right) {\bf L}_{ij}^{l l^{\prime}} }
\,\,\, V_{\rm eff} \left( \left| {\bf L}_{ij}^{ll'} \right| \right).   \label{eq2}
\end{eqnarray}
For this real space interaction model we use the simple form  
\begin{eqnarray}
\label{veff}
V_{\rm eff} \left(  d  \right) = 1 /  ( \epsilon_{r} \sqrt{ a_o^2 + d^2} ).
\end{eqnarray} 
Here $a_o$ accounts approximately for the reduction of Coulomb interaction strength 
at short distances due to $\sigma$ orbital polarization and 
delocalization of the $\pi$-charge density on each lattice site.\cite{zareagogolin} 
(In this equation energies are in Hartree ($e^2/a_{B}$) units and lengths are in units of the Bohr radius $a_{B}$.) 
In the real space model we choose the on-site interaction parameter $U$ separately from the longer 
range tail;
$U$ has been variously estimated as having values between $U\sim2 {\rm eV}$ to
$U\sim 6{\rm eV}$\cite{aliceayazyevbhowmickwunsch}, and up to an effective value of $U=9.3$ eV.
\cite{wehling} 
While Coulomb interaction energy at the carbon radius length scale is $\sim 20 {\rm eV}$,
and an estimate from the first ionization energy and electron affinity of a cabon atom
gives $U = 9.6 {\rm eV}$, \cite{dutta}
the effective on-site interaction strength is expected to be greatly reduced 
in the solid state environment because of screening by polarization of bound orbitals on 
nearby carbon atoms. 
For larger distance interactions we have included a factor $1/\epsilon_r$ to account for 
dielectric screening, as in the momentum space
version of the interaction model.
The value chosen for $\epsilon_r$ can be seen as an {\em ad-hoc} correction 
for overestimates of exchange interactions in Hartree-Fock theory.  
We study a range of values for this interaction parameter model 
but we believe that a value of $\epsilon_r \sim 4$ is normally appropriate for
graphene placed on a dielectric substrate. 
Values chosen for $\epsilon_r$ and $U$ control not only the overall
strength of the interaction term \cite{fuhrer} but also the 
relative strength of onsite and long range
parts of the interaction.
We will show later how this ratio can play a role in selecting the broken 
symmetry solutions which can appear in these models. 

\begin{figure}[htbp]
\begin{center}
\includegraphics[width=6cm,angle=90]{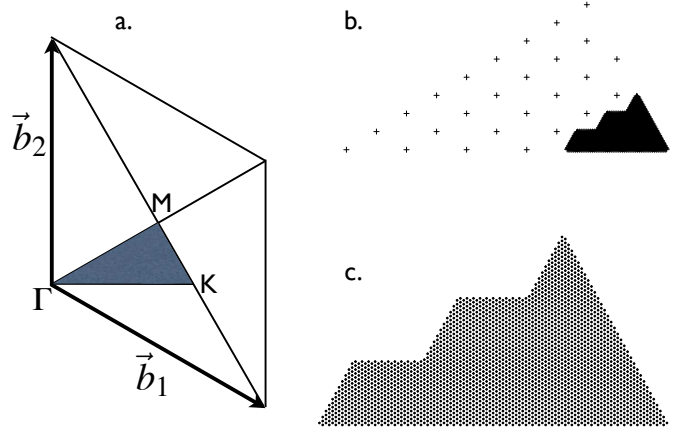} 
\caption{
Choice of the irreducible wedge of the primitive cell (left) and the adaptive sampling of the 
$k$-points in the vicinity of the Dirac point $K$ used for most of our calculations.
The density of $k$ points in the dense region shown in the figure corresponds to a sampling density 
of $512 \times 512$ points.
}
\label{adaptive}
\end{center}
\end{figure}

There are two technical difficulties in these calculations, one related to the nature 
of electron-electron interactions and one related to the electronic structure of graphene.
The long-range of the Coulomb interaction creates some numerical 
difficulties, particularly in evaluating the energies of the charge-density-wave states 
discussed below.   We have found the accurate results can be obtained by choosing a 
cut-off distance for the $1/r$ tail so that the coupled sites are as nearly as possible 
equally distributed between sublattices.  The second challenge is related to the 
band crossing at the Dirac point in graphene, at which the wavefunctions which enter 
the construction of the exchange potential have a singular dependence on wavevector.
Accurate calculations require dense $k$-point sampling
near the Dirac point, which increases the computational load rapidly 
in Hartree-Fock calculations because of the non-local exchange interactions.
In an effort to achieve a satisfactory compromise between computational load and accuracy we exploit
the hexagonal symmetry inherent in the problem. 
This allows us to limit our calculations to the irreducible wedge with $1 / 12$th of the Brillouin zone 
area, even thought the additional phase factors in the remainder of the zone need still to be properly 
accounted for when we calculate the exchange potential. 
We use denser adaptive $k$-point sampling 
near the Dirac cone while keeping a coarser grid in the remainder of the irreducible wedge as 
shown in Fig. \ref{adaptive}.
In this way it is possible to achieve good accuracy while maintaining the numerical load at a 
reasonable level. 
The coarse $k$ point sampling region was typically  kept to $16 \times 16$ density
while near the Dirac point we have chosen for most of our calculations 
a sampling density corresponding to $512 \times 512$ points in the full Brillouin 
zone and up to $1024 \times 1024$ density.

\section{Non-local exchange and logarithmic velocity divergence near the Dirac point}
Graphene's Dirac-like low-energy Hamiltonian\cite{novoselov,wallace}
provides an easily studied example of isolated band crossings near the 
Fermi level of a solid.
The band crossing at the two isolated Fermi points introduce 
singularities in the band Hamiltonian with interesting topological\cite{volovik} characteristics,
and facilitate the application of field-theoretic perturbative 
methods.\cite{logdivref}
As we will discuss later, there are some close analogies between interaction physics 
in graphene and in gapless\cite{gaplesssc} and narrow gap\cite{narrowgap}
semiconductors.
It has long been recognized that interaction effects can 
become prominent in gapless, semimetal, and narrow gap systems.
For example in a semimetal with a small overlap between valence and conduction bands, 
interactions can induce electron-hole pairing and turn the solid into an excitonic insulator.\cite{gaplesssc,halperin} 
In finite gap semiconductors Wannier-Mott excitons can form due to mutual attraction between a hole and an electron.
Non-local electron exchange interactions play a relevant role in defining the band structure of  
narrow band semiconductors. \cite{gelmont} 
In gapless semiconductors exchange induced corrections in the dispersion relation
are large near the crossing point and it has been argued that 
virtual generation of excitons can lead
to a dielectric anomaly.\cite{liu,gaplesssc}
A general study of materials with Fermi points has revealed that for linear band crossings,
interactions always introduce a logarithmically diverging
velocity enhancement,\cite{abrikosov} whereas 
instabilities are expected for quadratic crossings. \cite{abrikosov,abrikosov1}

\begin{figure}[htbp]
\begin{center}
\bigskip
\includegraphics[width=3.25cm,angle=270]{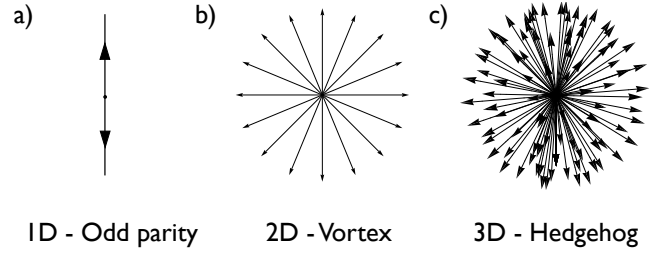} 
\caption{
Illustration of sublattice psedospin dependence on momentum for the Dirac-like Hamiltonians.
In 1D (a) the pseudospin changes direction at the Dirac point, in 2D (b) it has a vortex
and in 3D (c) it has a monopole hedgehog structure.
In each case the band state sublattice pseudospin changes direction 
upon crossing through the Dirac point.
}
\label{poles}
\end{center}
\end{figure}

\begin{figure}[htbp]
\begin{center}
\includegraphics[width=8.5cm]{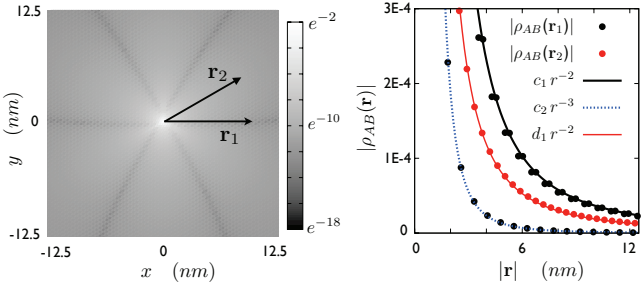} 
\caption{(Color online)
Absolute value of the sublattice off-diagonal density matrix 
defined in Eq. (\ref{offd}) illustrating the overall power law decay $r^{-2}$.
{\em Left Panel:} 
Grey scale map representation where we can notice a regular anisotropy 
of the off diagonal density matrix which follows the triangular Bravais lattice
structure of graphene.  The density matrix element falls off with a larger 
power law along certain discrete directions.  
{\em Right Panel:}
Density matrix at discrete lattice vectors along the directions
${\bf r}_1$ and ${\bf r}_2$ indicated in the left panel.
For direction ${\bf r}_1$ we notice a periodic dip in the value of
$\rho_{AB} ({\bf r})$ every three lattice constants.  The density matrix 
for these lattice vectors has a $r^{-3}$ decay law. 
The slow $r^{-2}$ power law decay reflects the 
singular dependence of valence band wavefunctions on ${\bf k}$ at the 
Dirac points.
The values of the fitting coefficients are $c_1 = 0.0037$, $c_2=0.0015$ and 
$d_1=0.0019$ when distances are measured in nm.
}
\label{offdiag}
\end{center}
\end{figure}

The marginal Fermi liquid behavior obtained in 3D,\cite{abrikosov} 
and in the graphene 2D case\cite{logdivref} is a consequence of non-local exchange interactions,\cite{logdivref1, logdivref2}
as we discuss at length below. 
To demonstrate explicitly how these velocity enhancements appear in our calculations we examine 
the Fock term in Eq. (\ref{hfgen}) expressed in the sublattice representation: 
\begin{eqnarray}
V_{X} \left( {\bf k} \right) &=& 
  \left( \begin{array}{cc}
  V_{X}^{ AA} \left( {\bf k} \right)  &
  V_{X}^{ AB} \left( {\bf k} \right) \\
  V_{X}^{ BA} \left( {\bf k} \right)  &  
  V_{X}^{ BB} \left( {\bf k} \right)
\end{array} \right).
\end{eqnarray}
The physics is most clearly explained using the real-space interaction 
version of our calculations, although the reciprocal-space version is 
more numerically convenient.
The diagonal matrix elements are identical by symmetry and
can be expressed using the real space sum of effective two body Coulomb 
repulsion in Eq. (\ref{eq2}). 
Using the symmetry property that 
 $ \left< c^{\dag}_{{\bf k}' A} c_{{\bf k}' A} \right> = \left< c^{\dag}_{{\bf k}' B} c_{{\bf k}' B} \right> = 1 / 2 $ 
for every value of ${\bf k}'$ in neutral graphene, we obtain
\begin{eqnarray}
V_{X}^{A\, A} \left( {\bf k} \right) &=&  
-  \frac{1}{2 N_{K}^2} \sum_{{\bf k}, i, \, j }^{N_{K}} e^{i \left(  {\bf k}' - {\bf k} \right) {\bf L}^{A \, A}_{i \, j}} 
V_{\rm eff} \left( \left| {\bf L}^{A \, A}_{i \, j} \right| \right) \\
&=&   - \frac{1}{2 N_{K}} \sum_{i \, j}^{N_{K}} \delta_{i,j} V_{\rm eff} \left( \left| {\bf L}^{A \, A}_{i\,j} \right| \right) = - \frac{U}{2}.
\end{eqnarray}
At half-filling, particle-hole symmetry implies that the sublattice-diagonal 
component of the density-matrix is half of the full $\pi$-band density-matrix, and therefore 
diagonal in lattice vector.  Only the on-site interaction contributes to
$V_{X}^{A\, A} \left( {\bf k} \right)$.  This contribution to the exchange energy is 
independent of momentum and does not contribute to the quasiparticle velocity.
For the off-diagonal term, on the other hand, we 
use the relation  $ \widetilde{\rho}_{AB} \left( {\bf k}' \right) = \left< c^{\dag}_{{\bf k}' B} c_{{\bf k}' A} \right> = - f \left( {\bf k}' \right) / 2 \left| f \left( {\bf k}' \right) \right|$
to obtain 
\begin{eqnarray}
\label{eq:vxab}
V_{X}^{A \, B} \left( {\bf k} \right) &=&  
  \frac{1}{2 N_{K}^2} \sum_{{\bf k}'} \sum_{ i, \, j }^{N_{K}} e^{i \left(  {\bf k}' - {\bf k} \right) {\bf L}^{A \, B}_{i \, j}} 
V_{\rm eff} \left( \left| {\bf L}^{A \, B}_{i \, j} \right| \right) \frac{f \left( {\bf k}' \right)}{\left| f \left( {\bf k}' \right) \right|}  \,\,\,\, \quad  \ \\
&\equiv& \frac{1}{2 N_{K}} \sum_{i, \, j}^{N_{K}} 
 e^{ - i  \, {\bf k} \, {\bf L}^{A \, B}_{i \, j}}
 \rho_{AB} \left( {\bf L}_{i \, j}^{A B} \right)  \,\, V_{\rm eff} \left( \left| {\bf L}^{A \, B}_{i \, j} \right| \right).   \label{eq:vxab1}
\end{eqnarray}
The second form for the right hand side expresses the exchange self-energy explicitly in terms of the 
sublattice off-diagonal element of the real-space density matrix: 
\begin{eqnarray}
\label{offd}
\rho_{AB} \left( {\bf L}_{i \, j}^{A \, B} \right) 
= \frac{1}{N_{K}} \sum_{{\bf k}^{\prime}} e^{i {\bf k}^{\prime}  {\bf L}_{i \, j}^{A B} }  
\widetilde{\rho}_{AB} \left( {\bf k}' \right).
\end{eqnarray}
In momentum space the Dirac band Hamiltonian's sublattice off-diagonal density matrix
is singular at the Dirac point because the valence band sublattice pseudospin state
changes at the Dirac point.  In a 1D model this effect leads to a discontinuity at the Dirac point, 
in 2D it leads to momentum space vortices, and in 3D to hedgehogs, as illustrated in Fig. \ref{poles}.
Because the function $f \left( {\bf k} \right)$ vanishes at the Dirac point, 
the inter-sublattice phase jumps along any line passing through it.
When this singularity is Fourier transformed to real space it leads to a slow 
power lay decay, as illustrated in Fig. \ref{offdiag} for the case of graphene, 
causing the electron exchange interaction to be strongly non-local.

The behavior of the real space tails can be obtained most simply from an analysis of the continuum model.
We redefine the wave vector ${\bf k}$ such that it represents the momentum measured from the Dirac point ${\bf K}$. 
A general three dimensional Hamiltonian with linear dispersion at an isolated band crossing
can be described by the Dirac-Weyl Hamiltonian
\begin{eqnarray}
H \left( {\bf k} \right) = \hbar \upsilon_{F} \sigma {\bf k} = 
\hbar \upsilon_{F}  k \left(
\begin{array}{cc}
\cos {\theta} & \sin {\theta} e^{-i \phi}  \\
\sin \theta e^{i \theta}  & - \cos \theta
\end{array} 
\right)
\end{eqnarray}
where $\sigma = \left( \sigma_x, \sigma_y, \sigma_z \right)$ is the Pauli matrix vector,
$k = \sqrt{k_x^2 + k_y^2 + k_z^2}$, $\tan \theta = \sqrt{ k_x^2 + k_y^2} / k_z$ and $\tan \phi = k_y / k_x$.
The density matrix for the occupied states is then given by
\begin{eqnarray}
\label{denmat}
\widetilde{\rho} \left( {\bf k} \right) = \frac{1}{2}\left(
\begin{array}{cc}
  \left( 1 - \cos \theta \right)  & -  \sin \theta e^{-i \phi}  \\
- \sin \theta e^{i \phi} &   \left( 1 + \cos \theta \right) \\
\end{array}
\right)
\end{eqnarray}
The 2D case is obtained by setting $\theta = \pi/2$ and 1D by setting $\phi = 0,\pi$.
For $ {\bf r} = {\bf L}_{i \, j}^{A \, B}$ in the $x$ direction, we
obtain the following result for the contribution to the density-matrix from a valley centered at ${\bf K}$: 
\begin{eqnarray}
\label{eq:rho}
\rho_{AB} \left( {\bf r} \right) &\simeq& 
e^{i{\bf K} \cdot {\bf r}}
\; \frac{A_0 }{\left( 2 \pi \right)^2} \int_{ \left| {\bf k} \right| < k_c} {\rm d} {\bf k} 
\,\,\, e^{i {\bf k}  {\bf r} }  \,\, \widetilde{\rho}_{AB} \left( {\bf k} \right)  \nonumber \\ 
&\propto& 
\left\{ 
\begin{array}{ll}
  \int_{-k_c}^{k_c} dk \,\, {\rm sgn}\left( k \right)  \exp(ikr)   & \quad 1D  \\
  \int^{k_c}_{0} dk \,\,  k \, J_{1} \left( k r \right)      & \quad 2D  \\
  \int_{0}^{\pi/2} d \theta \int_{0}^{k_c} dk  \sin^2 \theta \,\,\, k^2  \, J_{1} \left( k r  \sin \theta \right)    & \quad 3D  \\
 \end{array}
 \right.    \nonumber    \\
 &\simeq&   \frac{C_{d}}{r^d}
\end{eqnarray}
where $d$ is the dimension of the system and $J_1 \left( x \right)$ is a Bessel function of the first kind.
In graphene similar contributions are made by the two valleys.   The dominant 
contribution to this integral at large $r$ will come from the non-oscillatory $kr < 1 $ region 
when $J_1(x) \sim x/2$.  Inserting this limit into Eq. (~\ref{eq:rho}) and integrating 
up to $k \sim 1/r$ we see that $\rho_{AB} \left( {\bf r} \right) \sim r^{-d}$ at large $r$,
reminiscent of the dimensional dependence in the decay of Friedel oscillations.\cite{giuliani}
The off-diagonal density matrix in other directions differs only by a phase factor.

The slow power law decay behavior of the off-diagonal density matrix
in turn leads to a logarithmic divergence in 
$\nabla_{\bf k} V_{X}^{AB}({\bf k})$ evaluated using Eq. (~\ref{eq:vxab}) or Eq. (\ref{eq:vxab1}). 
We can obtain an approximate form for the exchange potentials in Eq. (\ref{eq:vxab1})
by changing the sum over discrete lattice sites to be a continuous integral
\begin{eqnarray}
V^{AB}_{X}\left( {\bf k} \right) 
&\simeq& \frac{1}{2 \Omega} \int d{\bf r}  \,\, e^{-i {\bf k} {\bf r}} \rho_{AB} \left( {\bf r} \right) V_{\rm eff} \left( \left| {\bf r} \right| \right)  
\end{eqnarray}
where $\Omega$ is the volume of the unit cell. 
Using polar coordinates to represent both ${\bf k}$ and ${\bf r}$ we evaluate the radial derivative 
of the exchange potential to obtain
\begin{eqnarray}
\frac{\partial V^{AB}_{X}\left( k \right)}{ \partial k }
&\simeq& \frac{1}{2 \Omega} \frac{\partial}{ \partial k} \int d{\bf r}  \,\, e^{-i {\bf k} {\bf r}} \rho_{AB} \left( {\bf r} \right) V_{\rm eff} \left( \left| {\bf r} \right| \right)  \nonumber \\
&=& \frac{-i}{2 \Omega}  \int d{\bf r}  \,\,  r \cos \widetilde{\theta}  \,\, e^{-i kr \cos \widetilde{\theta} } 
\rho_{AB} \left( {\bf r} \right) V_{\rm eff} \left( \left| {\bf r} \right| \right)  \,\,\,\,  \nonumber \\
&\simeq&   C_{\widetilde{\theta},d} \int_{a}^{k^{-1}} dr \,\, r^{d} \frac{1}{r^{d+1}}    \propto  {\rm ln} \left( \frac{1}{k a}\right),
 \label{radial}
\end{eqnarray}
where we integrated the angular variables first, identified 
the lattice constant as the lower limit of the approximate continuous position integral,
and $k^{-1}$ as the upper limit to avoid the oscillating regime.
Note that the space dimension drops out of the final result. 
Similar conclusions can be reached starting from Eq. (~\ref{eq:vxab}) 
and making a multipolar expansion of the Coulomb
interaction in momentum space.
\begin{figure}[htbp]
\begin{center}
\includegraphics[width=6.5cm,angle=90]{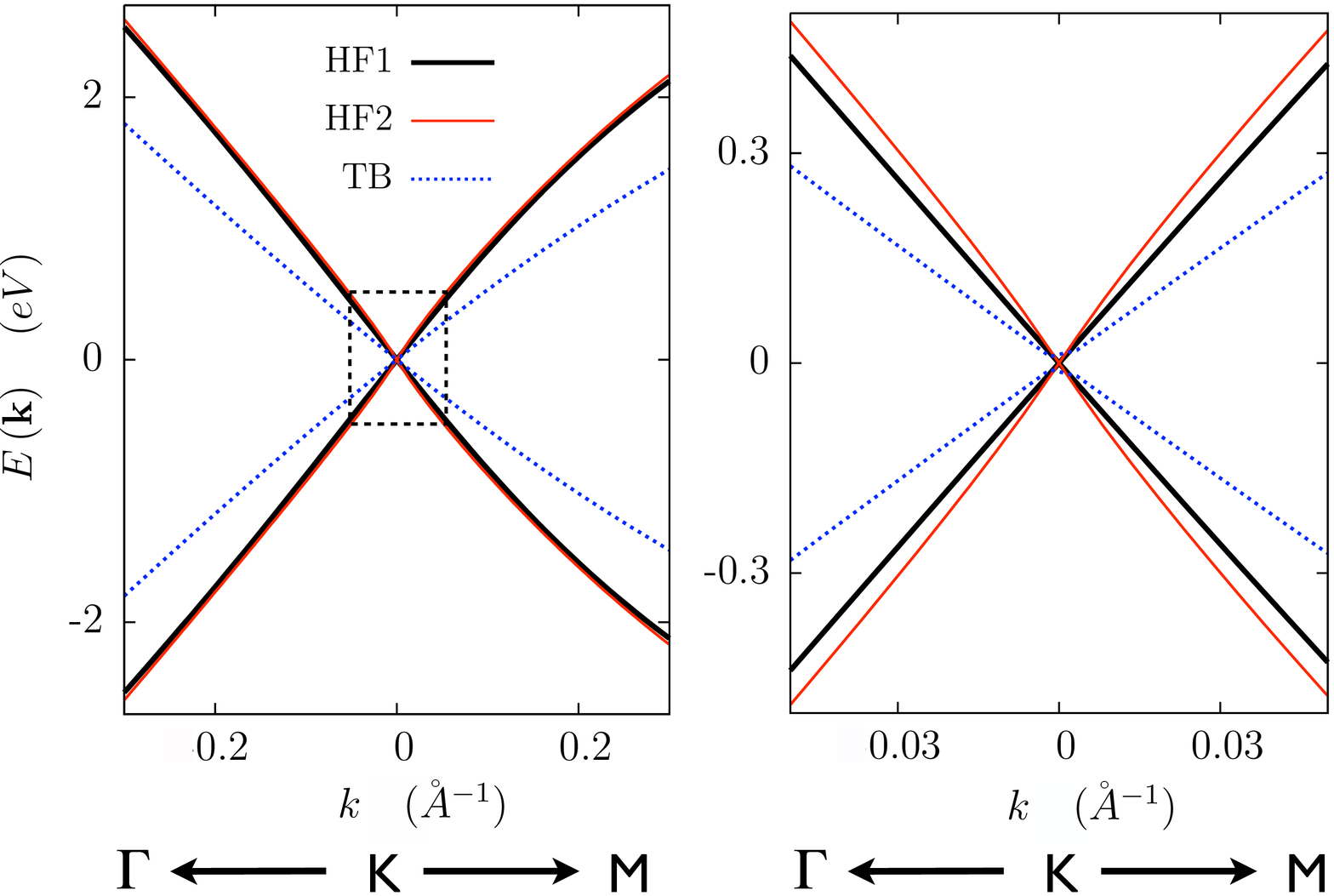} 
\includegraphics[width=6.5cm,angle=90]{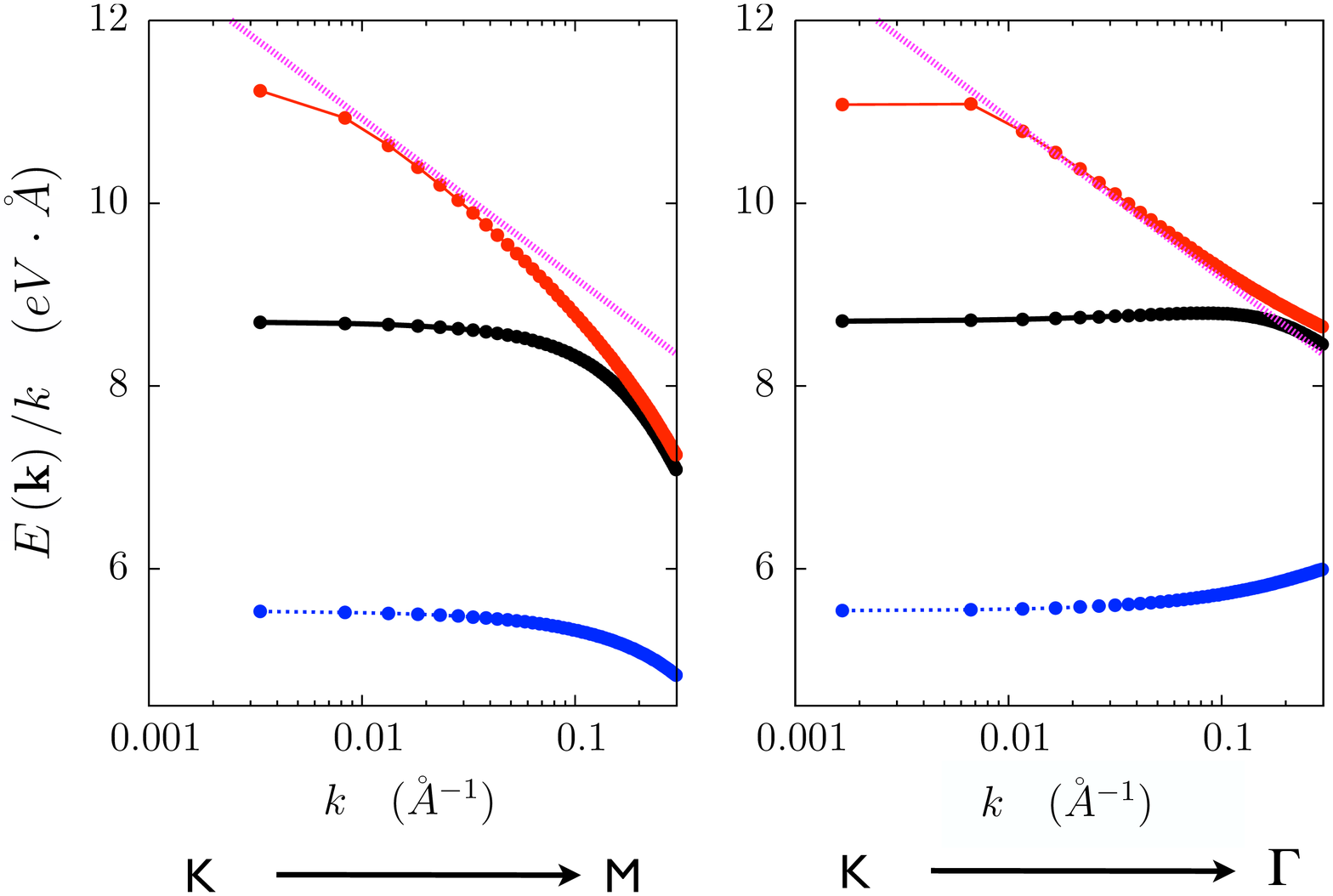} 
\caption{
{\em Upper panel:} 
Tight-binding and Hartree-Fock band structures near the Dirac point. 
The right panels blow up the the small rectangular regions shown 
in the left panels.
We observe that the momentum space Hartree-Fock calculation (HF2)
follows the enhancement to smaller momenta than the 
real space truncated interaction calculation (HF1).
{\em Lower panel:}
$E(k)/k$ versus $k$ close to the Dirac point.
The momentum space HF2 calculation used $1024 \times 1024$  $k$-points 
in the primitive cell.   The velocity enhancement saturates in both calculations.
The dashed straight line is the small $k$ fit obtained using Eq. (\ref{logcorr}) with $p_c = 30$.
}
\label{bstr}
\end{center}
\end{figure} 

In practical calculations both real-space and reciprocal-space Hartree Fock
calculations for graphene are able to 
follow the velocity enhancement only over a limited range of momenta,  
as illustrated in Fig. \ref{bstr}.  
The real space formulation used in the present calculation relies on a truncation of
the electron interaction range at about six lattice constants, as detailed in the appendix.
This prescription is able to describe a large part of the velocity increase due to 
non-local interactions, but saturates more quickly than the momentum space 
calculation which fails at small $k$ values due to the 
discreteness of the momentum sums used to construct the exchange Hamiltonian.

In Hartree-Fock continuum model calculations, the  
exchange-enhanced velocity is given by 
\begin{eqnarray}
\upsilon_{HF} = \upsilon_{F} \left(  1 + \frac{\alpha_{ee}}{4 } \, {\rm ln} \left(\frac{p_{c}}{k a} \right)  \right)
\label{logcorr}
\end{eqnarray}
where $\upsilon_{F} = \sqrt{3} \gamma_0 a / 2$ is the band velocity.  The logarithmic enhancement term 
has the prefactor $\alpha_{ee}/4$, where
$\alpha_{ee} = e^2/ \varepsilon_r \hbar \upsilon_F = (c/ \varepsilon_r \upsilon_F) \, \alpha$,   
is the effective fine structure constant,
$c$ is the speed of light, and $\alpha$ is the ordinary vacuum fine structure constant.
Our full Brillouin zone calculation allows us to obtain 
a numerical value for the dimensionless ultraviolet cutoff parameter $p_c$ 
in Eq. (~\ref{logcorr}). By fitting the numerical results we find that $p_{c} = 30 \pm 3$.

\section{Broken symmetry solutions phase diagram}
Recent lattice model Monte Carlo studies of interaction effects in graphene carried out by Drut and Lahde\cite{drutlahde}
predicted that they would be strong enough in suspended graphene samples to 
induce a CDW broken symmetry state with different electron densities on $A$ and $B$ 
and a gap emerges in the single-particle spectrum.
This broken symmetry in graphene is analogous to those that 
supply mass to elementary particles in particle physics.  
It now appears clear that these gaps do not occur in experimental samples,
possibly because of the role of lattice scale physics that is not 
reliably modeled in these simulations.
Indeed the size of the gaps must be fixed by ultraviolet physics because the 
two-dimensional Dirac model with Coulomb interaction does not define a
characteristic energy scale.  
The anticipated broken symmetries {\em do} occur in both lattice and continuum mean-field-theory 
models of single-layer graphene, although the interaction strengths at which they occur 
is likely underestimated by mean-field theory.  The calculations presented in this 
section demonstrate that the appearance or absence of these states is sensitive to 
lattice model detail,  in particular to the value of the on-site interaction
strength $U$ and the effective dielectric constant $\epsilon_r$.  
Studies of interactions based on Hubbard models
predict antiferromagnetic insulating states which appear for
$U \geq 2.23  \left| \gamma_0 \right| $ in Hartree-Fock mean-field-theory \cite{fujita,afqmc1} 
and for $U \geq \left( 4.5 \pm 0.5 \right)  \left| \gamma_0 \right|$ 
in Quantum Monte Carlo calculations.\cite{afqmc2}
A gapped spin-liquid state appears for $U\simeq 3.5 \left| \gamma_0 \right|$, \cite{meng}
before the AF state is reached, in the latter case.
In graphene, however, any attempt to estimate the character of the ground 
state must account for longer range interactions.\cite{herbut,wehling}

For the analysis carried out in this section we have used the real space formulation 
of the effective Coulomb interactions given in Eqs. (\ref{eq1}-\ref{veff}) that allows
a more direct control over the value of the onsite repulsion $U$ and the Coulomb interaction tail.
We used a model with finite truncation of the interaction range with a cutoff radius of about
six lattice constants. 
(Some considerations on optimal cutoff choices are 
explained in the appendix.)
Fig. \ref{phase} shows the mean-field phase diagram
produced by these calculations in which both spin-density-wave (SDW) and 
charge-density-wave (CDW) broken symmetry states appear.
\begin{figure}[htbp]
\begin{center}
\includegraphics[width=6.5cm,angle=90]{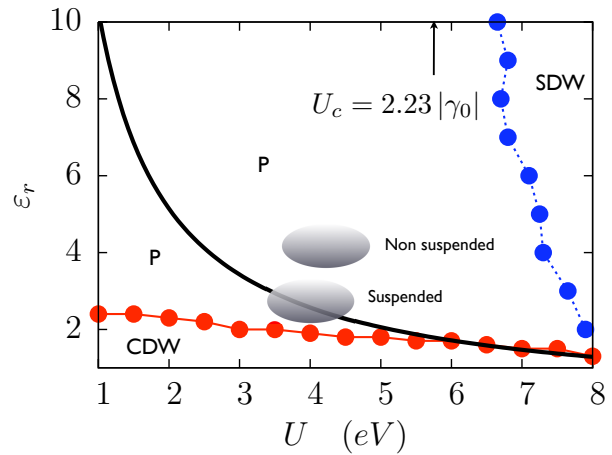} 
\caption{
Phase diagram showing where spin-density-wave (SDW) and 
charge-density-wave (CDW) broken symmetry solutions
appear in our model as a function of the interaction parameters $U$ and $\epsilon_r$.
Strong short distance repulsion (large $U$) favors SDW states, 
whereas weak short distance interactions and strong Coulomb interactions 
(small $\epsilon_r$) favors CDW states.  Below the the solid line in this figure 
the Hartree mean-field interaction energy is lowered by 
forming a CDW state which has different densities on A and B sublattices.
The CDW state boundary lies below this line because the band energy favors 
uniform densities.  The SDW state is a simple antiferromagnet, as expected at 
large $U$ on bipartite lattices.
The arrow in the figure shows the critical value
$U = 2.23 \left| \gamma_0 \right|$ beyond which SDW solutions appear for the pure 
Hubbard model.  The shaded regions in the figure indicate the parameter values 
thought to be most appropriate for graphene sheets that are suspended and for those 
that are supported by a dielectric substrate.
}
\label{phase}
\end{center}
\end{figure}
The solid line in the middle of the paramagnetic region of this figure follows 
$\epsilon_r \cdot U = 10.2838$ eV.  Along this line the Hartree mean field 
forming a charge density state with different densities on A and B sublattices 
vanishes.  The ordered states which appear above this line are spin-density-wave states,
which essentially reflect the physics expected for Hubbard models on a square lattice.
The ordered states which appear below this line are charge-density-wave states.
For large $U$ and small $\epsilon_r$ the charge-density-wave boundary is close to the 
the $\epsilon_r \cdot U = 10.2838$ eV  line, indicating that its location 
is determined mainly by this simple competition between short-range and long-range 
interactions.  When this consideration applies, 
CDW states cannot occur for $U > 10.2838$ eV since $\epsilon_r$ cannot take a value smaller than 1. 
A crude estimate of the onsite repulsion from the carbon atomic radius is $e^{2} / a_o \sim 20 \, $ eV 
whereas the value of $U$ that can be obtained from the first ionization potential and electron affinity of
carbon is $U \sim 9.6 \, $ eV. \cite{dutta}
The actual value will be further reduced when we account for additional screening effects from 
neighboring and onsite $\sigma$ orbitals, but the physically appropriate value 
is highly uncertain.
In our phase diagram CDW solutions, which are favored when the longer range part of the interaction
is strong but the short-range effective repulsion is weak, are restricted 
to values of $\epsilon_r \leq 2.2$ with small enough $U$.
  We conclude from this sensitivity that it is not possible to 
reliably predict the occurrence or absence of broken symmetry states on the basis of continuum model 
calculations alone.  The values of $\epsilon_r$ and $U$ thought to be appropriate based on 
considerations explained elsewhere \cite{aliceayazyevbhowmickwunsch,wehling}
are consistent with the absence of broken symmetry states in single-layer graphene samples.

\section{Discussion and conclusions}
In the present work we have presented a detailed analysis of mean field Hartree-Fock interaction 
effects in a lattice model of single-layer graphene.  
We first analyzed the velocity renormalization of the band dispersion near the Dirac point at 
the Hartree-Fock level.  These calculations demonstrate explicitly that the velocity 
enhancement is produced by non-local exchange interactions between different
graphene sublattices and provide a numerical estimate of a dimensionless ultraviolet parameter 
which cannot be estimated using Dirac continuum model calculations.  
Similar velocity renormalizations occur whenever a linear band crossing 
occurs at the Fermi level producing Fermi points.  In dimension $d$ the 
velocity enhancement is associated with a $r^{-d}$ power law decay in
the real space density matrix.  Large velocity enhancements will also 
occur for similar reasons whenever band gaps are small, or show semimetallic behavior
when the character of occupied states varies rapidly on the scale of the 
Brillouin-zone, although in this case they will always remain finite.  
This type of physics is responsible for strong the non-locality of exchange interaction 
in gapless or small gap semiconductors \cite{gaplesssc} with weak avoided crossing 
of the bands, in the surface states of topological insulators \cite{TI} 
or in metallic armchair carbon nanotubes. \cite{nanotube}

The velocity enhancements we explore in graphene are partially related to the Fermi surface enhancement 
incorrectly predicted by Hartree-Fock theory when it is applied to  
metals.\cite{giuliani}  In that case the enhancement is always suppressed by
screening.  In graphene, however, the density-of-states vanishes at the Fermi 
level and screening is less effective.  \cite{polini,sarma}
A random-phase-approximation theory 
which includes dynamic screening also predicts logarithmic enhancement of 
the velocity, but with a slightly modified logarithm prefactor.  

Our mean field study of broken symmetry states is summarized by the phase diagram as a function of Coulomb 
interaction parameters in Fig. \ref{phase}.  We have shown that 
CDW states are favored by weak on-site interactions and 
or SDW states by strong on-site interactions, but that neither instability 
occurs in a broad range of interaction parameter space.  
The most realistic values for the two parameters are still not accurately known, 
but may be guessed from the character of the broken symmetry states 
which do in fact occur in the quantum Hall regime of graphene
in which the kinetic energy is quenched \cite{bfield}.
Our suggested values for these parameters, both for suspended and unsuspended 
samples are shown in Fig. \ref{phase}.
According to the phase diagram we have obtained,     
suspended samples of graphene without substrate
dielectric screening ($\epsilon_r \sim 1$) is likely reasonably 
close to a CDW instability.
This result is in rough agreement with the lattice Monte-Carlo calculations of Drut-Lahde \cite{drutlahde}
who predict a band-gap opening for graphene for a critical value of $\epsilon_r \sim 1$.
However, the latest available transport measurements for suspended 
graphene \cite{bolotin} find a finite resistivity of about 16$k\Omega$ in agreement with early 
predictions\cite{ando} for the minimum conductivity for graphene.
There is no experimental evidence for an insulating CDW state.
This discrepancy between experiment and present theory signals in 
part the limitations of $\pi$-band only models
that do not include screening of the bare electron
by carbon $\sigma$ band polarization.
An increase of the effective dielectric constant 
from $\epsilon_r=1$ to $\epsilon_r \sim 2$ to account for screening by
degrees-of-freedom not included in the $\pi$-band model would be sufficient to 
explain the absence of broken symmetry states in suspended samples.
Recent inelastic X-ray scattering experiments\cite{abbamonte} in graphite 
find screening at high energies within graphene sheet.
These results motivate further efforts to estimate high-energy screening
in monolayer graphene.

{\em Acknowledgments.}  
We gratefully acknowledge helpful discussions with Dima Pesin.
Financial support was received from Welch Foundation grant TBF1473, 
NRI-SWAN, and DOE grant DE-FG03-02ER45958
from the Division of Materials Sciences and Engineering .

\appendix
\section{Real space truncation of the Coulomb interaction}

We discuss below the optimum choice for the real-space interaction cutoff.
Even though the definition of effective Coulomb integrals in real space 
has a physically transparent meaning, one important drawback is that
the long range of the Coulomb repulsion makes 
sums over lattice sites of Eqs. (\ref{eq1}) and (\ref{eq2}) 
have slow convergence. 
A simpler method than the more accurate Ewald sum \cite{ewald}
consists in introducing a finite spherical truncation of the 
electron interaction range \cite{spherical} 
as an extended Hubbard model where we incorporate farther neighbor 
contributions in the Coulomb term.
For many purposes this method yields correct enough answers
because the effective reach of the Coulomb interaction shrinks 
when the positive background charge is taken into account.
Because of the slower decay in real space of the direct 
Coulomb term compared to the exchange potential
the inaccuracy in the electrostatic energy is usually the largest source 
of error of this truncation method specially when there is no charge neutrality 
within the interaction cutoff range in presence of inhomogeneous density distributions. 
One way to minimize this error is to choose the cutoff range such that
the electrostatic energy is minimized in presence of an symmetric charge imbalance 
in the A and B sublattices of graphene.
\begin{figure}[htbp]
\begin{center}
\includegraphics[width=6.5cm,angle=90]{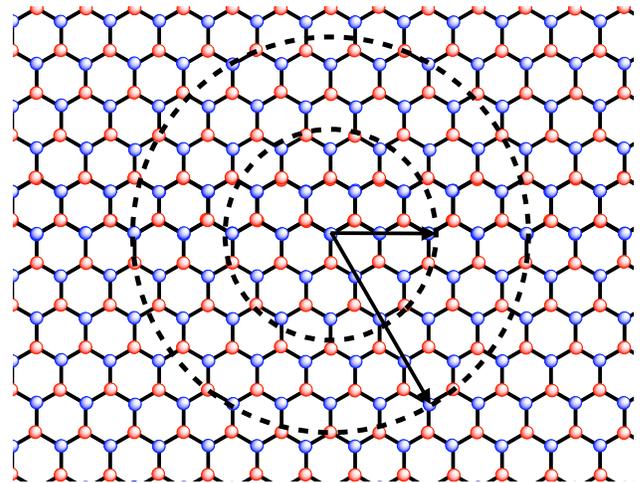} 
\caption{
Real space truncation of the interaction range in graphene as illustrated with two different
cutoff values of $L_{max} \sim 2a$ and $L_{max} \sim 5a$.
As we change the value of the cutoff radius $L_{max}$ there are oscillations in the relative number of 
carbon lattice sites $A$ and $B$ enclosed within the cutoff distance.
}
\label{default}
\end{center}
\end{figure}
In order to evaluate the cutoff for the Coulomb interaction term that minimizes
the error we express the Hartree energy of a CDW state
\begin{eqnarray}
E_{H} &=& \frac{1}{2} \int {\rm d}{\bf r} {\rm d}{\bf r}^{\prime} 
\frac{n\left( {\bf r} \right) n \left( {\bf r}^{\prime} \right) }{ \left| {\bf r} - {\bf r}^{\prime} \right|}
\simeq \frac{1}{2} \sum_{i,j} n_{i} n_{j} V_{ij}
\end{eqnarray}
where we use the notation $V_{ij} = V_{\rm eff} \left( d_{ij} \right)$ for simplicity where
$d_{ij}$ is the distance between the lattice sites $i$ and $j$.
Let us consider a charge density transfer of $\delta n$ from lattice B to lattice A
such that the densities are $n_{A} = n_0 + \delta n$ and $n_{B} = n_0 - \delta n$.
In that case we obtain
\begin{eqnarray}
E_{H}  
&=& \frac{1}{2} \sum_{i \in A, j \in A} V_{ij} \left( n_0 + \delta n \right) \left( n_0 + \delta n \right) \\
&+&  \frac{1}{2} \sum_{i \in B, j \in B} V_{ij} \left( n_0 - \delta n \right) \left( n_0 - \delta n \right) \\
&+& \frac{1}{2} \sum_{i \in A, j \in B} V_{ij} \left( n_0 + \delta n \right) \left( n_0 - \delta n \right) \\
&+& \frac{1}{2} \sum_{i \in B, j \in A} V_{ij} \left( n_0 - \delta n \right) \left( n_0 + \delta n \right).
\end{eqnarray}
\begin{figure}[htbp]
\begin{center}
\includegraphics[width=6.5cm,angle=90]{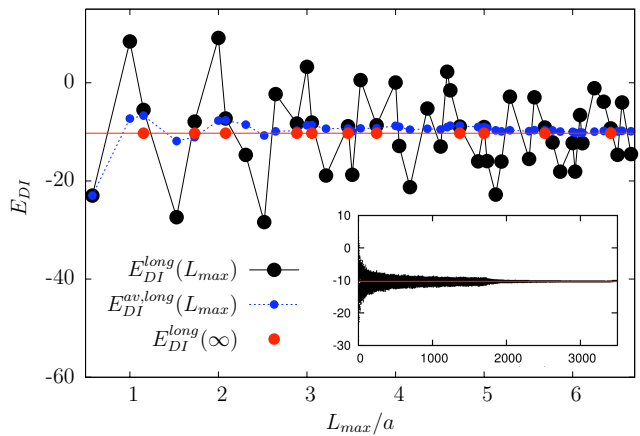} 
\caption{
Longer ranged contributions to the Hartree energy $E^{long}_{DI} \left( L_{max}  \right)$
as defined in Eq. (\ref{longerrange}) which shows a strong cutoff distance $L_{max}$
dependent oscillation that converges slowly to the limiting value represented
with the horizontal line whose behavior is more clearly shown in the inset. 
We can notice that $E^{long}_{DI} \left( L_{max}  \right)$ is rather close to the asymptotic
limit for certain values of $L_{max}$.
A better estimate of the asymptotic limit can be obtained from the behavior of
$E^{av, long}_{DI} \left( L_{max} \right)$ defined in the text.
}
\label{harcut}
\end{center}
\end{figure}
The linear terms in $\delta n$ above cancel each other 
and if we neglect a constant shift in the origin 
the electrostatic energy difference per lattice is
\begin{eqnarray} 
\label{eq:di} 
\delta E_{DI} 
&=& \frac{\left( \delta n \right)^2}{2} \; \Big[ U + \sum_{j \in A}   
 V_{ij} - \sum_{ j \in B} V_{ij} \Big] 
\end{eqnarray} 
where $d_{ij}$ is the distance between lattice sites $i$ and $j$, 
$U = V(d_{ii})$ and $i$ is a fixed label belonging to sublattice $A$. 
We denote the cutoff dependent direct energy corresponding to the long 
ranged part of the Coulomb interaction as 
\begin{eqnarray}
E^{long}_{DI} \left( L_{max} \right) &=& \sum_{j \in A, \, L_{max}} V_{ij} - \sum_{ j \in B, \,L _{max}} V_{ij}
\label{longerrange}
\end{eqnarray}
which shows an oscillatory dependence on the cutoff distance $L_{max} > d_{ij}$ 
as represented in Fig. \ref{harcut}.
This behavior poses some caveats in extended Hubbard models with only one or two 
neighbor Coulomb interactions when used for obtaining a phase diagram of broken 
symmetry states involving charge density modulations or comparing results between
different models. 
We can clearly observe that the above mentioned oscillations slowly converge 
to a constant for very large $L_{max}$.
A better estimate for the asymptotic value in the limit $L_{max} \rightarrow \infty$ can be obtained from 
$E^{av, long}_{DI} \left( L_{max} \right) = \sum_{i=1}^{N_{max}} 
E^{long}_{DI} \left( L_{max, i} \right) / N_{max}$
averaging the values obtained at each discrete $i^{th}$ nearest neighbor shell cutoff, where $N_{max}$ is
the total number of nearest neighbor shells corresponding to the cutoff distance $L_{max}$.
We can observe that for certain specific values of $L_{max}$ the quantity
$E^{long}_{DI} \left(L_{max} \right)$ is close to $E^{long}_{DI} \left( \infty \right)$.
In table I we represent the values of some of these select cutoff distances 
which are the ones that minimize the difference in the number of A and B lattices 
and therefore minimizes the deviation from charge neutrality for a CDW state within 
the cutoff range. 
In our calculations we have used a cutoff just above the value $L_{max} = 6.4291 a$ listed in the table.
\begin{table}[ht] 
\caption{
Optimum values of cutoff $L_{max}$ and the corresponding values of $E^{DI}_{long} \left( L_{max} \right)$
which gives the closest estimates to the asymptotic limit $E^{DI}_{long} \left( \infty \right)$ for each 
period of oscillation. For completeness we also represent $E^{av}_{DI}\left( L_{max} \right)$ defined in the text.
We denote with the superscript 1 the results obtained with $a_o = a/\left( 2 \sqrt{3} \right)$ in the definition 
of the effective Coulomb integral in Eq. (\ref{veff}) and superscript 2 the results we would obtain if we used $a_o = 0$.
 } 
\centering 
\begin{ruledtabular}
\begin{tabular}{c|cc|cc} 
$L_{max}$   &     $E^{long,1}_{DI}$   & $E^{av, 1}_{DI}$  &  $E^{long,2}_{DI}$   & $E^{av, 2}_{DI}$  \\
\hline  
       1.1547     &   -5.5346       &  -6.7019  &-10.5026     &   -12.0711   \\
       1.7321     &   -7.9043       &  -11.0792  &-13.2176     &   -16.5852 \\ 
       2.0817     &   -7.2731       &  -7.6482  &-12.5287     &   -13.0159\\
       2.8868     &   -8.2996       &  -9.7492  &-13.6225     &   -15.1359 \\
       3.0551     &   -8.0970       &  -8.6223   &-13.4115     &   -13.9862  \\
       3.4641     &   -8.8583       &  -9.3227    &-14.1986     &   -14.6905    \\
       3.7859     &   -8.6589       &  -9.2623   &-13.9943     &   -14.6260  \\
       4.7258     &   -8.9231       &  -8.7407    &-14.2648     &   -14.0952  \\
       5.0000     &   -9.0172       &  -9.0100    &-14.3603     &   -14.3650 \\ 
       5.6862     &   -9.1180       &  -9.6429     &-14.4631     &   -14.9990   \\
       6.4291     &   -9.3049       &  -9.8026    &-14.6521     &    -15.1579 \\
       7.0946     &   -9.5278       &  -9.8566    &-14.8769     &    -15.2111  \\
       7.3711      &   -9.4074       &  -9.5712   &-14.7556     &   -14.9247  \\
       8.0829     &    -9.5431       &  -9.6167    &-14.8922     &   -14.9698  \\
       8.3267     &    -9.4921       &  -9.6134    &-14.8409     &  -14.9662  \\ 
       8.7369     &     -9.6973      &  -9.9986   &-15.0472     &   -15.3518  \\ 
       9.8150     &     -9.7037      &  -9.8903    &-15.0536     &  -15.2428   \\
      10.0167    &      -9.6323     &   -9.6926  &-14.9819    &    -15.0447  \\
      10.1489    &     -9.6717      &  -9.7987    &-15.0215    &    -15.1509   \\
      10.4403    &     -9.7986      &  -9.9958    &-15.1488    &   -15.3481  \\
      10.6927    &     -9.7210      &  -9.8310   &-15.0709    &   -15.1831  \\
      11.6762     &     -9.7431      &  -9.8290    &-15.0931    &   -15.1809  \\
      11.8462    &     -9.7469       & -9.8273    &-15.0970     &   -15.1791 \\
      12.2202      &    -9.8575     &  -10.0860  &-15.2078    &    -15.4379 \\
      12.4231     &     -9.8141     &  -10.0185  &-15.1643    &   -15.3703  \\ 
      12.5033    &      -9.8081     &   -9.9754    &-15.1583    &   -15.3272  \\
      \hline 
\multicolumn{1}{c|} {$\infty$}               & 
\multicolumn{2}{c|} {-10.2838 }                       &
\multicolumn{2}{c} { -15.6327 }     \\
\end{tabular} 
\end{ruledtabular}
\label{table:nonlin}  
\end{table}

\end{document}